\documentclass[twocolumn,aps,pre,showpacs,amsmath,amsfonts,amssymb,floatfix]{revtex4}
\usepackage{morefloats}
\usepackage{amsmath}
\usepackage{graphicx}
\usepackage{dcolumn}
\usepackage{bm}
\usepackage{color}
\definecolor{orange}{rgb}{1,0.5,0}
\begin{document}
\title{Impact of leader on cluster synchronization}

\author{$^{1, 2}$Sarika Jalan\footnote{sarika@iiti.ac.in} }
\author{$^1$Aradhana Singh }
\author{$^3$Suman Acharya}
\author{J\"{u}rgen Kurths$^{4, 5}$ } 
\affiliation{$^1$Complex Systems Lab, Indian Institute of Technology Indore, Indore-452017}
\affiliation{$^2$Centre for Bio-Science and Bio-Medical Engineering, Indian Institute of Technology Indore, Indore-452017}
\affiliation{$^3$ Physical Research Laboratory, Navrangpura, Ahmedabad-380009}
\affiliation{$^4$ Potsdam Institute for Climate Impact Research, D-14412 Potsdam }
\affiliation{$^5$ Institute for Complex Systems and Mathematical Biology, University of Aberdeen,
Aberdeen-AB243FX}

\begin {abstract}
We study the mechanisms of frequency synchronized cluster formation in
coupled non-identical oscillators and investigate the impact of presence of
a leader on the cluster synchronization. We find that the introduction 
of a leader, node having large parameter mismatch, induces  
a profound change in the cluster pattern as well as in
the mechanism of the cluster formation. The emergence of a leader generates a 
transition from the driven to the mixed cluster state. The frequency mismatch turns out to be 
responsible for this transition. Additionally,
for a chaotic evolution, the driven mechanism stands as a primary mechanism for the cluster formation, whereas for a periodic 
evolution the self-organization mechanism becomes equally responsible. 
\end{abstract}
\maketitle

The interaction between individual units of a system leads to many 
emerging behaviors, among which
synchronization is one of the most fascinating phenomenon which has been
gaining tremendous attention since the first experimental demonstration of the
phenomena by Christian Huygens \cite{Huygens}.
The unexpected sway and twist of the Millennium bridge,  synchronization of the neurons in the brain and synchronous fire flies \cite{Strogatz_bridge,book_syn,book_neuron} are few examples of the synchronization in the real world systems.
Synchronization is defined as an emergence of some relation between the 
functional of two processes due to interaction \cite{book_syn}.  
 Earlier studies on coupled 
non-identical oscillators have shown that the exact synchronization is hard to 
achieve when there is a parameter mismatch in the local dynamics
\cite{book_syn}, rather they exhibit phase or generalized synchronization \cite{Kurths, Freq_syn}. 
Additionally, there exists a nontrivial transition to the global
phase synchronization in a population of globally 
coupled chaotic non-identical oscillators \cite{Glob_oscillator}. 
 Moreover, cluster pattern and frequency of the nodes in a cluster  have been shown to be controlled by local external forcing as well as by changing the network architecture \cite{Scholl}. 
Furthermore, the coupled oscillators with heterogeneous coupling have been reported to exhibit synchronization triggered by the oscillators having strong couplings,  further facilitating the synchronization among the nodes having weak coupling \cite{Hetero_coup}.

We present results of cluster synchronization due to its
importance and occurrence in various
real world systems represented in terms of interacting units \cite{Clus_syn}. 
We study mechanisms of formation of frequency synchronized clusters in coupled 
non-identical oscillators and investigate the influence of a leader on the dynamical evolution of other nodes. 
One possible way of
defining a leader is to make the natural frequency of a node much higher than that
of other nodes in the network \cite{Leader_high_freq,Lead_follow}. This is one of the traditional way to define a leader in the coupled dynamics on network models. 
Some other ways of defining
leaders are those which depend upon the application and motivation of the problem, e. g., Ref.\cite{Leader} considers a leader which exchanges information with its neighbors as well as has access to its own state.
In Ref.\cite{Neuron_leader} the neuron which fires first is considered as leader.
Further, a leader can be assigned based on its degree in a network \cite{Network_leader}. Our work
reveals 
that difference of the natural frequency of a node with the rest
decides its impact on the cluster synchronization and dynamical evolution of all other nodes.
We present the results for the coupled R\"ossler oscillators on various possible
networks such as 1-d lattice, scale-free and random networks. 
Earlier works have shown that the network properties, such as degree and betweenness centrality, play an important role in the synchronizability of coupled oscillators \cite{betweenness_syn1},
and based on the analysis of small-world networks it has been shown that the synchronizability can be 
enhanced by raising the maximum degree of the network
as well as by reducing the maximum betweenness \cite{betweenness_syn2}. 
A recent work on power-grid also emphasizes on the structural importance of perturbed nodes for stability of synchronized state \cite{powergrid_kurths}.

This Rapid Communication reports that
a combination of the degree and natural frequency mismatch of a node with other nodes decides the role
of a leader in the network, particularly, its impact on the phenomena of cluster formation. We demonstrate that, an enhancement in the betweenness centrality does not enhance the impact of a leader if the degree
is maintained.
Apart from the impact of a leader on the cluster synchronization, we report various different mechanisms of cluster formation in the presence of a leader. 
 The earlier works on the coupled maps have identified two different phenomena for the cluster synchronization, namely, the driven (D) and the self-organized (SO) \cite{SJ_prl2003}.
The SO synchronization refers to the state when clusters are formed due to the intra-cluster couplings and D synchronization  refers to the state when clusters are formed due to the inter-cluster couplings. Furthermore, for coupled chaotic oscillators having randomly distributed frequencies, it has been reported that with an increase in the coupling strength, the nodes with smaller frequency mismatch synchronize and form a synchronized cluster \cite{Kurths}, while we find that if the natural frequency of the nodes are distributed in a narrow band, at lower couplings the synchronization between pair of nodes which are not directly connected is preferred. Using a simple star network model, we demonstrate that the D mechanism is preferred over SO due to a common coupling environment D synchronized nodes may have. 
Further, presence of a leader causes an enhancement in the synchronization between the directly connected nodes thus leading to a transition from the D to the mixed or dominant SO clusters state.
We study phase synchronization of the coupled oscillators, where the dynamics of the $i$-th oscillator can be written as:
\begin{equation}
\dot{X}_i = f(X_i,\omega_i) + \frac{\varepsilon}{k_i} \sum_{i=1}^N A_{ij}  h(X_i, X_j); \; i = 1, \ldots, N
\label{dyn_jth_osc}
\end{equation}
where, $X_i \in R^m$ is $m$ dimensional state vector of the $i$-th oscillator and $f: R^m \rightarrow R^m$ provides the dynamics of an isolated oscillator and $h$ is the coupling function. 
$A$ is the adjacency matrix of the network defined as: $A_{ij}= 1$, if oscillators $i$ and $j$ interact, otherwise $A_{ij}=0$. The degree of a node is defined as $ k_{i}$ = $\sum_{j=1}^{N}A_{ij}$  and the parameter $\varepsilon$ defines the strength of overall coupling among the oscillators.

Coupled R\"ossler oscillators on networks can be written as a set of the following ordinary differential equations:
\begin{eqnarray}
& & \dot{x_i} = -\omega_{i}y_{i} - z_{i}  \nonumber \\
&  & \dot{y_i} = \omega_{i}x_{i} + ay_{i} + \frac{\varepsilon}{k_i}\sum_{j=1}^N A_{ij} (y_j - y_i)  \nonumber \\
& & \dot{z_i} = b + z_{i}(x_{i}-c)
\label{Rossler_eq}
\end{eqnarray}
Here $\omega_{i}$ is the natural frequency of the $i^{th}$ oscillator which we consider randomly distributed in the interval $1< \omega< 1.05$ \cite{Kurths}. We take a node acting as a leader when its natural frequency
 is much greater than the rest of the nodes in the network ($\omega_L \gg 1.05$).  Later on we will explain that the strength of this natural frequency mismatch of a node together with the degree of the node decide the impact of a leader in a network.  
In Eq.~\ref{Rossler_eq}, $a=0.15$, $b=0.4$ and $c=8.5$ for which the uncoupled dynamics is chaotic \cite{book_syn}. 
Further the phase $\theta$ and the averaged partial frequency of the $i$-th oscillator can be defined as $
\theta_i = \arctan\frac{y_i}{x_i} $ and  $\Omega_i  =  <\dot{\theta_i}(t)>$ respectively.
 Here, we note that the frequency $\Omega_i$ is in general different from the intrinsic frequency $\omega_j$.
The uncoupled oscillators (i.e. $\varepsilon =0$ in Eq.~(\ref{dyn_jth_osc})) evolve independently. 
 With an increase in $\varepsilon$, the formation of synchronized clusters is observed as $\varepsilon$ exceeds a critical value $\varepsilon_{cs}$. Frequency of oscillators may be synchronized forming clusters, 
i.e. $\Omega^l_i = \Omega^l; \; j=1,\ldots,N_l$, where $\Omega^l$ is the synchronization frequency of cluster $l$ and $N_l$ is the number of oscillators in the $l$-th cluster and 
$l=1,\ldots,C$; $C$ is the maximum number of clusters. 

\begin{figure} 
\centerline {\includegraphics[width=2.17in, height=1.2in]{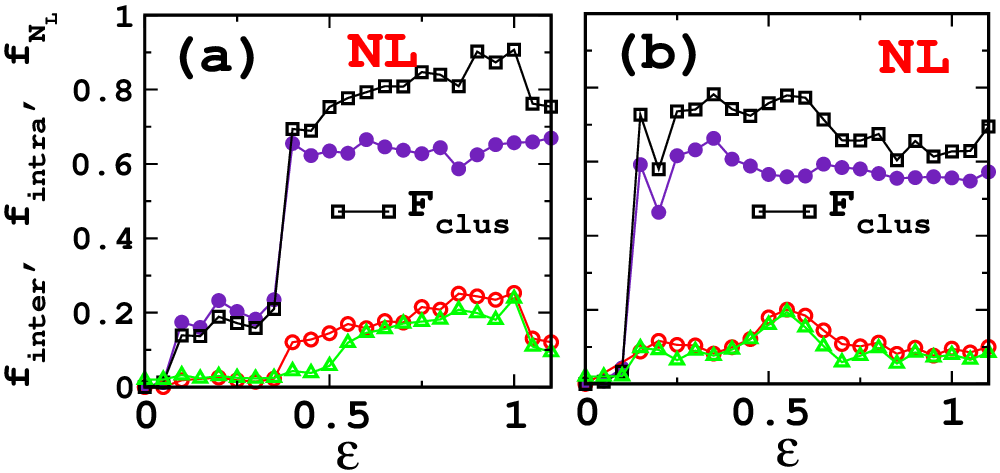}
\includegraphics[width=1.17in, height=1.18in]{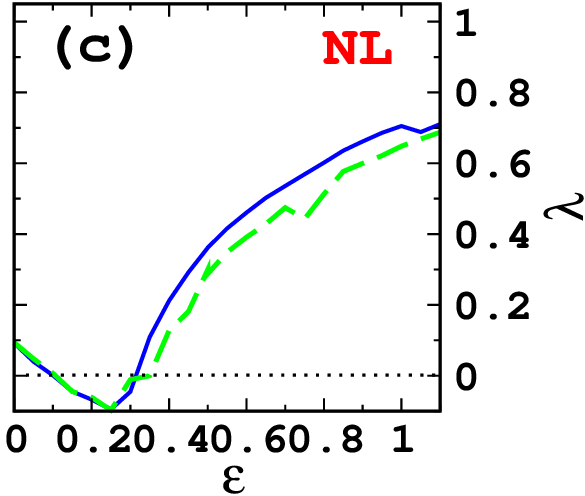}}
\centerline{\includegraphics[width=3.25in, height=1.18in]{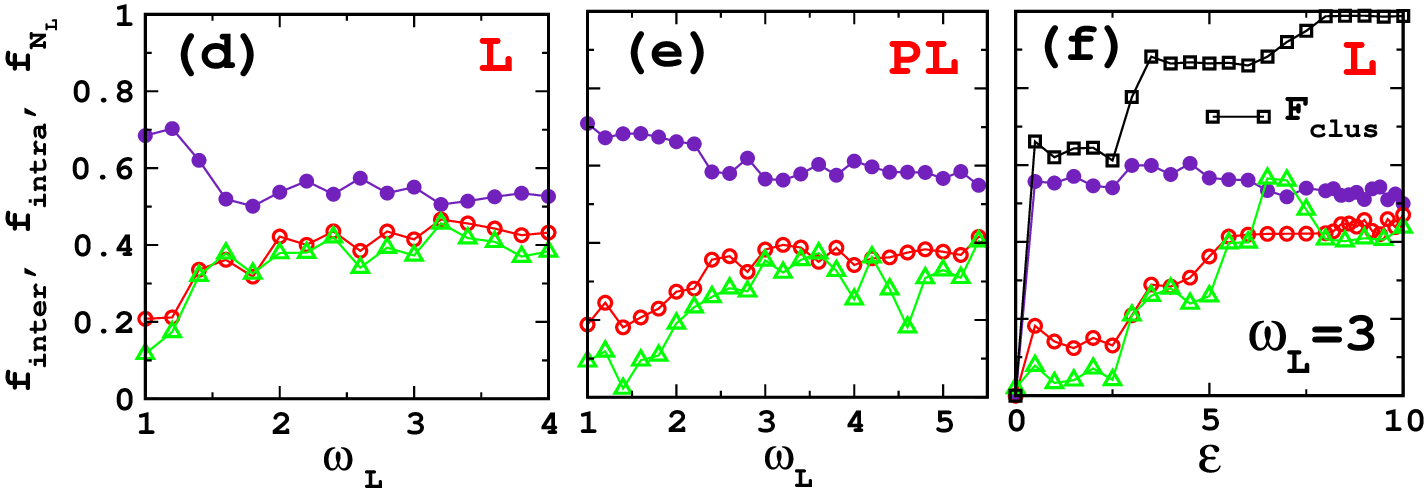}}
\caption{(Color online) (a) and (b) present variation of $f_{inter}$ (closed circle), $f_{intra}$ (open circle), $f_{N_L}$ (fraction of nodes in the largest cluster)(open square) and $F_{clus}$ (fraction of nodes forming clusters) (open triangle)  with increase in $\varepsilon$ and $\omega_L$ for the random (a) and scale-free networks (b), (d), (e), (f) of $N=50$, $\langle k \rangle =2$. (c) presents the variation of the largest Lyapunov exponent ($\lambda$) with change in the coupling strength for the random (solid line) and scale-free (dashed line) networks. 
$N L$, $L$ and $P L$ represent no-leader hub being leader and peripheral node being leader case respectively. The figures are obtained by taking average over 20 random initial conditions.}
\label{Fig_Rand_SF_leader}
\end{figure}

Depending on the connections between the nodes, represented by the adjacency matrix, and the synchronized clusters, three  phenomena of cluster 
formation have been identified ; driven(D), self-organized(SO), and mixed \cite{SJ_prl2003}.
The quantities $f_{intra} = N_{intra}/{N_c}$ and $f_{inter} = N_{inter}/{N_c}$, stand as measures for 
SO and D clusters respectively,
where $N_{intra}$ and $N_{inter}$ are the numbers 
of intra- and inter- cluster couplings, respectively. In 
$N_{inter}$, couplings between two isolated nodes are not included.
${N_c}$ is the total number of connections in the network.
The state which correspond to $f_{intra} \cong 0$ and $f_{inter} > 0$ is defined as the ideal D clusters state; $f_{intra} >0$ and $f_{inter} = 0$ correspond to the ideal SO state; 
$f_{intra} \neq 0$ and $f_{inter} \ggg f_{intra}$ correspond to the dominant D and 
$f_{inter} \neq 0$ and $f_{intra} \ggg f_{inter}$ correspond to the dominant SO clusters state. We take $\lvert f_{intra} - f_{intra} \rvert <0.2$, to define mixed clusters state \cite{mixed_clus}. For the higher values, dominant D and dominant SO region 
will shrink and mixed region will grow, while for lower values reverse will happen. 
Further, we define a cluster pattern as a particular state which contains
information of all the pairs of synchronized nodes distributed in various
clusters in the network. A change in the pattern refers to the case when the nodes in the different clusters get rearranged \cite{PRE_2013}.
Furthermore, we define cluster synchronizability in terms of number of the nodes participating in a cluster. Based on this, we say cluster synchronizability enhances if the number of nodes participating in the clusters increases. Further, there might be a situation when all the nodes participate in cluster, for that cluster synchronizability may be enhanced if the size of a cluster increases or the total number of the clusters reduces. 

Starting with a set of random initial conditions, we evolve the coupled dynamics (Eq.\ref{Rossler_eq}) on 
different networks, namely, 1-d lattice, scale-free, and random networks. After an initial transient we study the cluster synchronization. We consider the evolution of coupled oscillators without any leader, followed by the investigation of synchronized clusters
in the presence of a leader. In the following we discuss the cluster synchronization for all the networks. 

The heterogeneity in degree of the
scale-free networks \cite{Scale-free_net} provides several options for choosing a leader in the network yielding very different structural properties to the leader.
For example, a hub may be assigned as a leader making it the highest degree node
and consequently best connected with rest of the nodes in the network, or a periphery node may be assigned as a leader which makes
the leader worse connected with the rest of the nodes. 
\begin{figure}[t]
\begin{center}
\includegraphics[width=0.7\columnwidth]{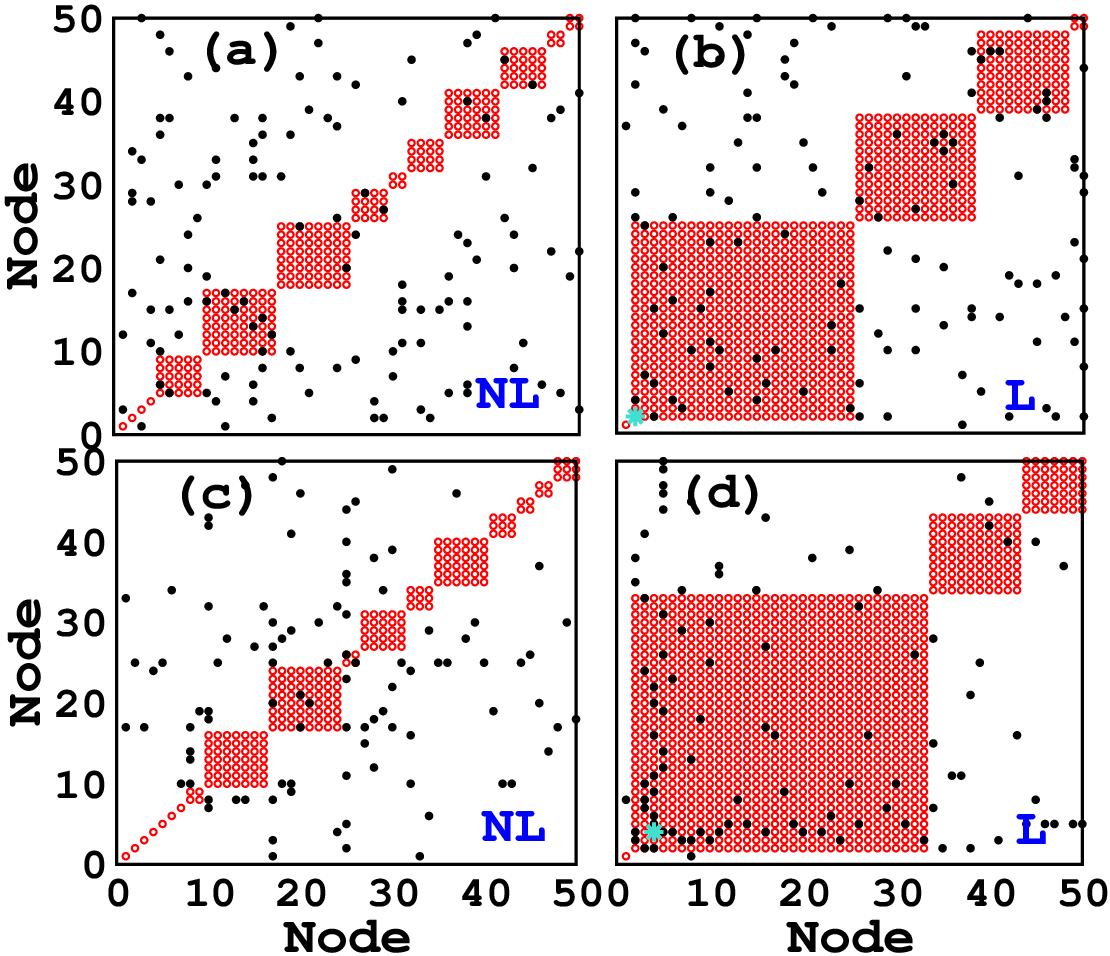}
\end{center}
\caption{(Color online) Snap shots showing the change in the cluster pattern for random and scale-free networks of $N=50$, $\langle k \rangle =2$ at $\varepsilon=8$. Figs (a), (d) are plotted for random network and Figs (c), (d) are plotted for scale-free network. The open(red) dots show that the corresponding nodes are synchronized and the closed(black) dots show that the corresponding nodes are connected. The snap shot is plotted after renumbering the nodes so that the nodes forming a cluster come nearby. The star represents the position of the leader in the cluster after rearranging the nodes.
$N L$ and $L$ represent no-leader and  leader case respectively.}
\label{Fig_snapshot}
\end{figure}
As depicted in Fig.~\ref{Fig_Rand_SF_leader}(a), without 
a leader the coupled dynamics (\ref{Rossler_eq}) leads to the dominant D clusters at all the couplings except at very high values where mixed clusters exist.
For sparse networks considered here, we find that while the network exhibits a 
good  
cluster synchronization, accompanied with many small clusters, the maximum number of nodes in a cluster does not exceed
$20\%$ of the network size. The D clusters  correspond to the chaotic 
evolution as reflected by the largest Lyapunov exponent 
(Fig.~\ref{Fig_Rand_SF_leader}(c)). 
What follows that a small mismatch in the natural 
frequencies of the directly connected nodes does not allow them to 
synchronize with each other, whereas the synchronization between a pair
of nodes which are connected through other nodes gets facilitated through a common 
coupling environment. 
For example, the frequency of the non-identical peripheral nodes in a star network synchronize with each other, while leaving hub out of the clusters \cite{remote_kurths} as the dynamics of the peripheral nodes from Eq.\ref{dyn_jth_osc} can be written as:
\begin{equation}
\dot{X}_i = f(X_i,\omega_i) + \varepsilon (X_h-X_i); \; i = 2, \ldots, N
\nonumber
\end{equation}
The hub provides the common coupling to the peripheral nodes and thus drive them to form a D synchronized cluster. 
It is however interesting to observe the similar behavior for other sparse networks, which consist of many star like structures instead of having the ideal situation.

Upon making a node as a leader by enhancing its natural frequency higher than that of the other nodes, we find that the coupled dynamics leads to a transition from the dominant D to the mixed clusters state (Fig.~\ref{Fig_Rand_SF_leader}(f)).
The natural frequency of the leader, which leads to this transition depends on the degree of the node.A hub node being the leader generates the transition at relatively lesser frequency as compared to that required for a peripheral node being the leader. For example, for a hub being a leader, the mixed cluster state is observed for $\omega_L \gtrsim 2$ (Fig.~\ref{Fig_Rand_SF_leader}(d)),  while for the peripheral node being the leader the mixed clusters are observed for $\omega_L \gtrsim 3$ (Fig.~\ref{Fig_Rand_SF_leader}(e)). 

\begin{figure}[t] 
\includegraphics[width=2.8in, height=1.0in]{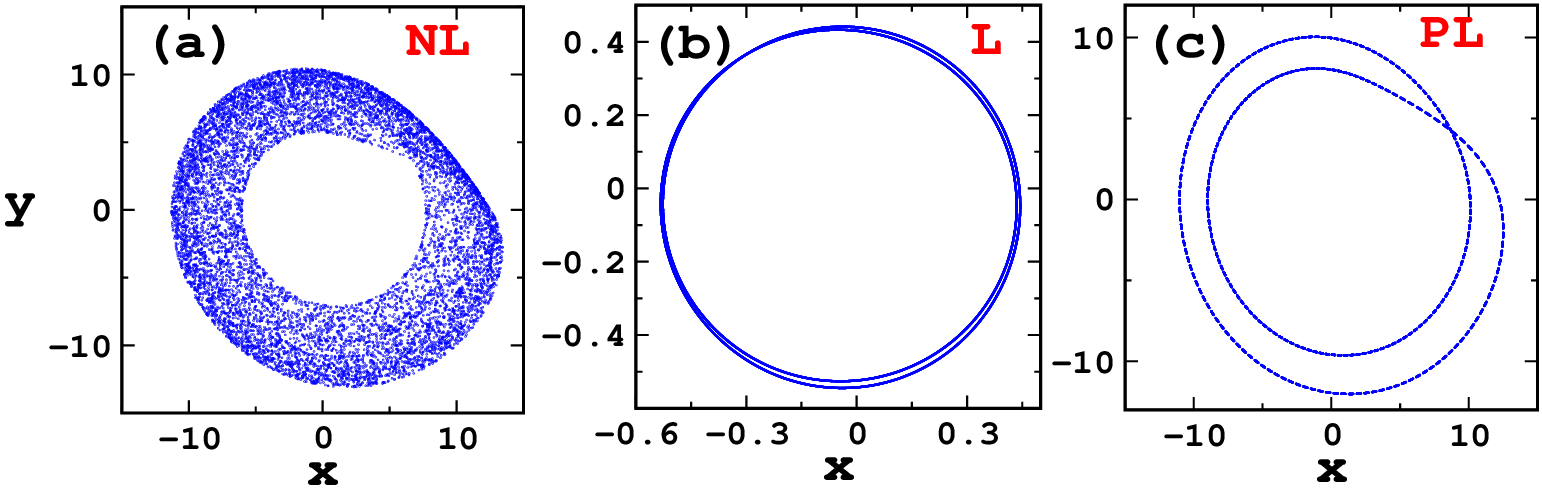}
\caption{(Color online) Projection of the phase portraits of coupled R\"{o}ssler oscillators for the no-leader and leader case for a scale-free network of $N=50$, $\langle k \rangle =2$ at $\varepsilon=2.4$.}
\label{Fig_xy}
\end{figure}

For a hub being the leader, at weak couplings ($\varepsilon < 2.2$) the clusters 
remain to be governed by the D mechanism (Fig.~\ref{Fig_Rand_SF_leader}(c)) as observed for the without leader. 
With an increase in the coupling strength, for $ \varepsilon >3$, there is an enhancement in
the SO synchronization. We emphasize that as the D mechanism is still playing a role in the cluster formation, with the
inclusion of the SO mechanism, the final cluster state becomes of the mixed type.  Moreover, number of nodes in the largest cluster increases (Fig.~\ref{Fig_Rand_SF_leader}(f)). Additionally, in 
the same coupling regime there is a change in the dynamical 
evolution. The dynamics in this regime becomes periodic (Fig.~\ref{Fig_xy}(b) and (c)) against the chaotic evolution
observed for the no-leader case (Fig.~\ref{Fig_xy}(a)). 

Another impact of the inclusion of a leader in the network is that
it may lead to a completely different clusters pattern (Fig.~\ref{Fig_snapshot}(b) and (d)) than observed for the no-leader case (Fig.~\ref{Fig_snapshot}(a) and (c)). 
 
Inclusion of a peripheral leader  at small couplings
leave the dynamical evolution unchanged.
Whereas at strong couplings the frequency mismatch of the nodes connected with the leader 
enhances the SO synchronization similar as observed for the hub being the leader. 
This enhancement in the
SO synchronization can be explained using the revelation that the parameter mismatch between two
nodes leads to a more stable synchronization \cite{REA_2014}, where node having
large natural frequency dominates the evolution with
other nodes which are directly connected with it leading to the synchronization.

\begin{figure} 
\centerline{\includegraphics[width=0.8\columnwidth]{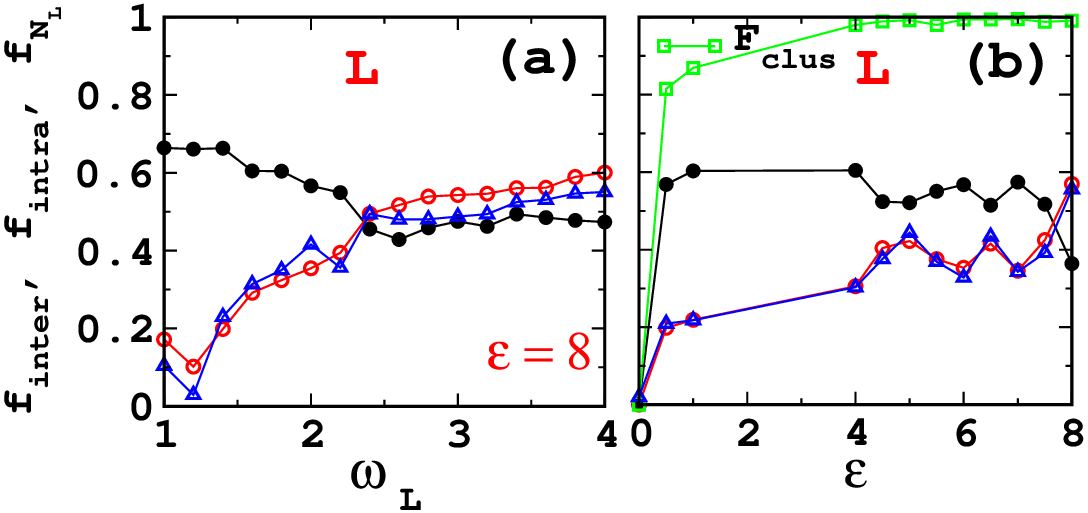}}
\centerline{\includegraphics[width=0.8\columnwidth]{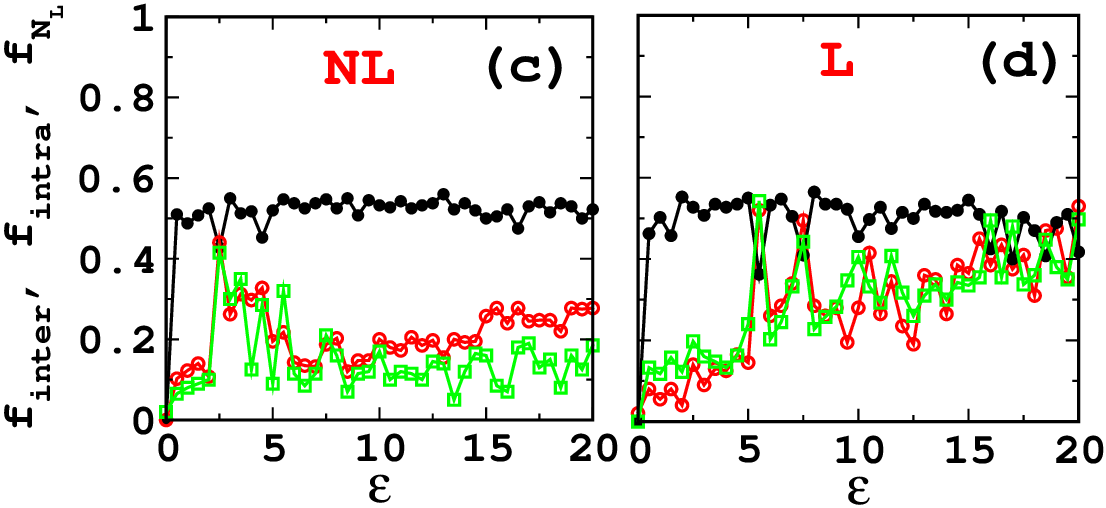}}
\caption{(Color online)  $f_{inter}$, $f_{intra}$ and $f_{N_L}$ as a function of  $\omega_L$ and $\varepsilon$. (a), (b) correspond to the random network and (c), (d) correspond to the 1-d lattice of $\langle k \rangle =4$ and $N=50$. All graphs are plotted for average over 20 realizations of the initial conditions.
}
\label{Fig_Rand_NNC_finter}
\end{figure}

We remark that the impact of a leader, whose degree lie in between
the highest and the lowest degree, lies in between these two. 
For example, presence of a leader in the ER networks \cite{Scale-free_net} the coupling range $0 \lesssim \varepsilon \lesssim 4.0$ keeps the driven phenomena behind the synchronization intact, however there is an increase in the SO phenomena.
With a further increase in the coupling strength, for $\varepsilon > 4.0 $, the SO synchronization enhances further as indicated by the enhancement in the value of $f_{intra}$ leading to the mixed clusters (Fig.~\ref{Fig_Rand_NNC_finter}(b)). The enhancement in the SO synchronization is associated with the enhancement in the fraction of nodes in the largest cluster as depicted in the Fig.~\ref{Fig_Rand_NNC_finter}(b).
The snap-shots in Fig.~\ref{Fig_snapshot} depict that inclusion of the leader decreases the number of clusters while keep almost all the nodes forming the clusters, thereby enhancing the cluster synchronization as defined in the model section. 

The 1-d lattice provides an ideal example of a homogeneous network.
The introduction of a leader in this arrangement leads to a similar behavior as exhibited
by the random and scale-free network depending upon the degree of the lattice which also becomes the degree of the leader (Fig.~\ref{Fig_Rand_NNC_finter}(d)), which
on one hand rules out any impact of structural
position or preference of the leader and on other hand demonstrates importance
of the degree of the leader in a network. What follows that the impact of a leader on 
other nodes is high if
the leader has a large degree, in the absence of which the leader should increase the coupling
strength in order to bring upon the same impact.

\begin{figure}[t] 
\centerline{\includegraphics[width=0.8\columnwidth]{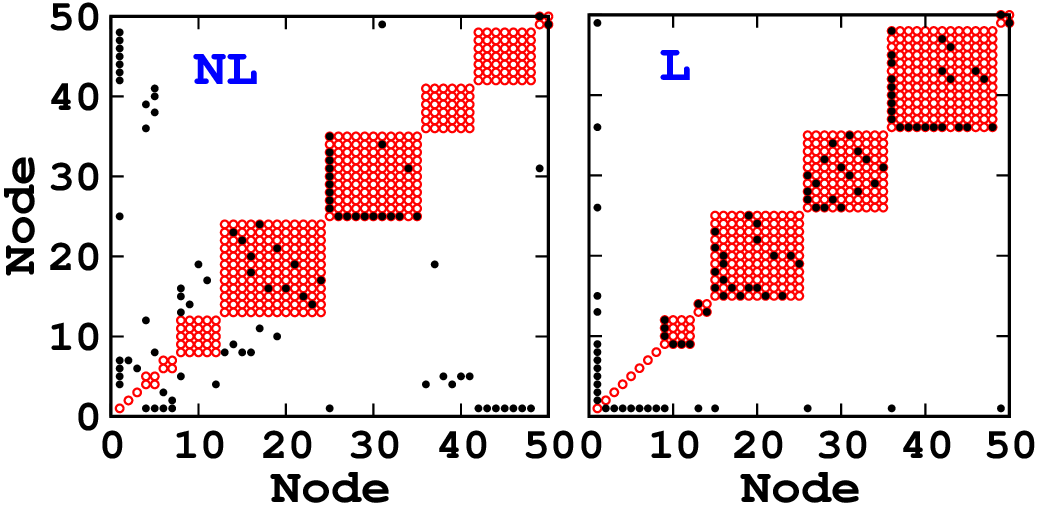}}
\caption{(Color online) Snap shots depicting changes in the cluster pattern for scale-free networks with $N=50$, $\langle k \rangle =2$ at $\varepsilon=3.2$. The natural frequency of the nodes are randomly distributed in the interval $1<\omega <1.09$ and $\omega_L = 10$.
}
\label{Fig_Kuramoto}
\end{figure}

 Further, in order to demonstrate the robustness of the phenomenon that the introduction of a leader exhibits a change in the mechanism of cluster formation, we consider Kuramoto model which is one of the most celebrated mathematical model for coupled oscillators \cite{Kuramoto}.
\begin{equation}
\dot{\theta_i}(t) = \omega_i + \frac{\varepsilon}{N}\sum_{i=1}^N A_{ij} sin(\theta_j - \theta_i) .
\end{equation}
We find that introduction of a leader leads to a transition from the dominant D to the  dominant SO clusters state. Fig.~\ref{Fig_Kuramoto} demonstrates formation of the mixed or ideal D clusters without a leader and SO cluster in the presence of a leader. 

To conclude, we investigate the cluster synchronization and phenomena behind the cluster formation for the diffusively coupled R\"{o}ssler oscillators and find that 
 a leader with its natural frequency much higher than that of other nodes, has a significant impact on the cluster synchronization. The cluster synchronizability of the network is enhanced either through an  enhancement in the number of the nodes participating in the cluster formation or due to a merging of several smaller clusters in to larger clusters or due to the formation of larger clusters consisting of a completely new set of nodes.
Further investigations reveal that the introduction of a leader may lead to a transition from the D to SO mechanism of the cluster formation.
 Thus, in the presence of a leader, synchronization between the directly connected nodes is enhanced. The presence of a leader may also lead to a transition from the chaotic to the periodic dynamical evolution and hence a leader may be introduced for chaos control \cite{book_syn} by tuning frequency of a single node. 
Interestingly, a leader has the maximum impact on the cluster synchronization if it is the highest degree node in the network rather than being the node which has the highest betweenness centrality. If a leader has a lower degree, its natural frequency should be relatively higher in order to achieve the enhancement in the cluster synchronization. For homogeneous networks, where all the nodes have the same degree,  coupling should be high in order to have a transition from the dominant D to the mixed cluster state. 
Furthermore, the presence of a leader not only changes the phenomena behind the cluster formation but may also completely change the cluster pattern. 


Leaders naturally arise in real-world networks such as in social 
networks \cite{Leader_social_book}, neural networks \cite{Neuron_leader}, 
protein translation regulatory networks \cite{Nature_ribosome}.  
In social network, a leader may possess one of the characteristics 
of power, experience, fame or wealth, while in biological networks, 
such as in  neural and the protein translation regulatory networks, 
a leader may be one which initiate certain processes \cite{Neuron_leader, 
Nature_ribosome}. 
Our work may be extended to capture particular properties of a 
leader for understanding the origin of synchronized clusters 
in these systems \cite{Clus_syn}. 
For example, a leader may have different coupling strength, 
such as in the brain network, where the synapses become 
weak with age \cite{book_neuron,Brain}.

{\it Acknowledgement:} 
SJ acknowledges DST (SR/FTP/PS-067/2011) and CSIR (25(0205)/12/EMR-II) for the financial 
support. SA acknowledges Complex Systems Lab for providing conducive atmosphere during the work.


\begin{thebibliography}{99}
\bibitem{mixed_clus} The value of threshold considered here enables us in identifying the transition from the dominant D to the mixed cluster state. A lower value of the threshold leads to diminution of the mixed regime and expansion of the dominant D regime, as a consequence upon inclusion of the leader, the same value of $f_{inter}$ and $f_{intra} $ which brings the transition from D cluster state to the mixed cluster state, will indicate a dominant D clusters state only. Similarly higher values of  the threshold will expand the mixed cluster region, and will shrink the dominant D region, thus without inclusion of a leader the same value of $ f_{inter}$ and $f_{intra}$ will indicate the mixed clusters and hence again we will fail to witness any transition. Furthermore, a change in the threshold will affect the boundaries of the dominant SO cluster state, but we do not observe any new phenomenon based on shifting this boundary.
\bibitem{Scale-free_net} The scale-free networks are constructed by starting with 
$\langle k \rangle$ nodes and then adding one node with $\langle k \rangle$ connections 
at each step \cite{rev_network}. The Erd\"os-Renyi random networks are constructed by making connections between all the pairs of
the nodes with the probability $p$ \cite{rev_network}.
\end{thebibliography}
\end{document}